\documentclass[preprint]{elsarticle}
\usepackage{graphicx}
\usepackage{epsfig}
\usepackage{amssymb} 
\usepackage{amsmath}
\usepackage{amsbsy} 
\usepackage{color}

\def\Frac#1#2{\frac{\displaystyle #1}{\displaystyle #2}}

\begin{document}

\title{Influence of reflector materials and core coolant on the characteristics of accelerator driven systems}

\author[infn]{F. Panza}
\ead{fabio.panza@ge.infn.it}

\author[infn]{M. Osipenko}
\ead{osipenko@ge.infn.it}

\author[infn,centrofermi]{G. Ricco}
\ead{ricco@ge.infn.it}

\author[infn,centrofermi]{M. Ripani}
\ead{marco.ripani@ge.infn.it}

\author[infn]{P. Saracco\corref{cor1}}
\ead{paolo.saracco@ge.infn.it}

\cortext[cor1]{Corresponding author}

\fntext[infn]{I.N.F.N. - Sezione di Genova, V. Dodecaneso, 33 - 16146 Genova (Italy)}
\fntext[centrofermi]{CENTRO FERMI - Compendio del Viminale  -€"  Piazza del Viminale 1 -€" 00184 Rome (Italy)}



\begin{abstract}
In this paper we simulated the behavior of a simple ADS model, based on MOX fuel embedded in solid lead, in terms of multiplication coefficient $k_{eff}$, thermal power and absolute neutron spectra.  In the first part of the paper, we report on the results obtained when modifying the reflector surrounding the fission core, by replacing pure lead with a layered graphite/lead structure. We found that, by appropriately choosing position and thickness of the graphite and lead layers, it is possible to obtain a "hybrid" system where the neutron spectrum in the core still exhibits a fast character, while the spectrum in the graphite layer is considerably softer, becoming thermal in the most peripheral positions. In order to obtain such a modulation of the neutron spectra from the center of the system to the periphery, a careful choice of the materials has to be made in order to avoid large variations of the local power at the core boundary. However, the smoothness of the power distribution is obtained at the expense of lower values of $k_{eff}$ and the total power of the system. In the second part of the paper, we explored the option of adopting light water as coolant, instead of the helium gas assumed in the initial design. We found that this produces an increase in $k_{eff}$ and thermal power, without significantly perturbing the fast character of the system and without introducing spatial power excursions in any place within the core. The characteristics obtained may allow to design a system where fast, mixed and thermal spectra can be used to expand the use of the ADS as an irradiation facility.
\end{abstract}

\maketitle

\vfill
\noindent

\section{Introduction}
Nuclear systems are divided into two main categories: thermal and fast. In the first case the presence of a light material (i.e. light or heavy water, graphite) is necessary as a moderator, able to slow down the fast neutrons from fission reactions in order to increase the fuel fission cross section. This way a large thermal power can be produced with a moderate enrichment of the fuel in fissile nuclides content, but due to the equally large capture cross section, long-lived radioactive wastes are produced as well by a sequence of neutron captures and $\beta$ decays starting from $^{238}U$, which leads to the production of Plutonium and Minor Actinides (MA).

Instead, fast reactors don't use any moderator, in order to minimize the slowing down of neutrons, and use a liquid metal (sodium, lead) or gas as coolant.  An important advantage of this system is that for fast neutrons the fission cross section in $^{238}U$ becomes larger than capture, so that less heavy long-lived nuclides are produced. Moreover, since many of the MA undergo fission by fast neutrons, they can also be burnt up in the reactor core itself. All this can result in a reduced production of long-lived radioactive waste \cite{GenIV}.

In this paper we studied the effects of the use of moderator materials in the design of fast nuclear systems, analyzing the corresponding issues as well as the possible advantages from a purely physical point of view. Our goal was to explore the feasibility of a low-power, relatively low-cost hybrid system, where by hybrid we mean that both hard and soft spectra arise in different zones, assuming that such a facility could be interesting for well-defined experimental, demonstration, educational and training purposes, in contrast with large scale facilities for irradiation purposes, like, e.g., the Jules Horowitz reactor \cite{JHR}. 

For our studies we assumed a simplified ADS \cite{Nifene} system, inspired by previous work and based on a solid lead matrix \cite{RiccoEd}, with power low enough to still keep the lead in a solid state and to present less demanding safety aspects.

In Section 1 we describe and simulate such test installation composed by a MOX-fueled and He-cooled ADS, that in the rest of the paper will be identified as the "initial model".

Starting from this model, in Section 2 a composite lead/graphite reflector was investigated, while in Section 3 the effects of light water as coolant replacing helium were examined. Conclusions are summarized in Section 4. The characteristics obtained may be used as the basis to design an irradiation facility where fast, mixed and thermal spectra can be present at the same time.

\section{The model}
In order to investigate the effect of moderator materials in a fast nuclear system, we start from a schematic subcritical research reactor (ADS) model of the lead type (Fig. 1a), composed by a cylindrical core with a 50 cm radius and 90 cm height, surrounded by a lead reflector with 120 cm radius and 160 cm height.  Core and reflector are contained within a 5 cm thick AISI steel vessel. As neutron source driving the core we assumed (p,n) reactions produced by 70 MeV protons fully absorbed by a Beryllium target, placed inside a 6 cm radius central axial beam pipe filled with helium gas as coolant. For a 1 mA proton current, the produced neutron intensity is about $8\,10^{14}$ neutrons/sec \cite{Misha}, however in the following all computed observables will be given per single source neutron. except for power distributions.
\begin{figure}[!htb]
\centering
\includegraphics[width=\linewidth]{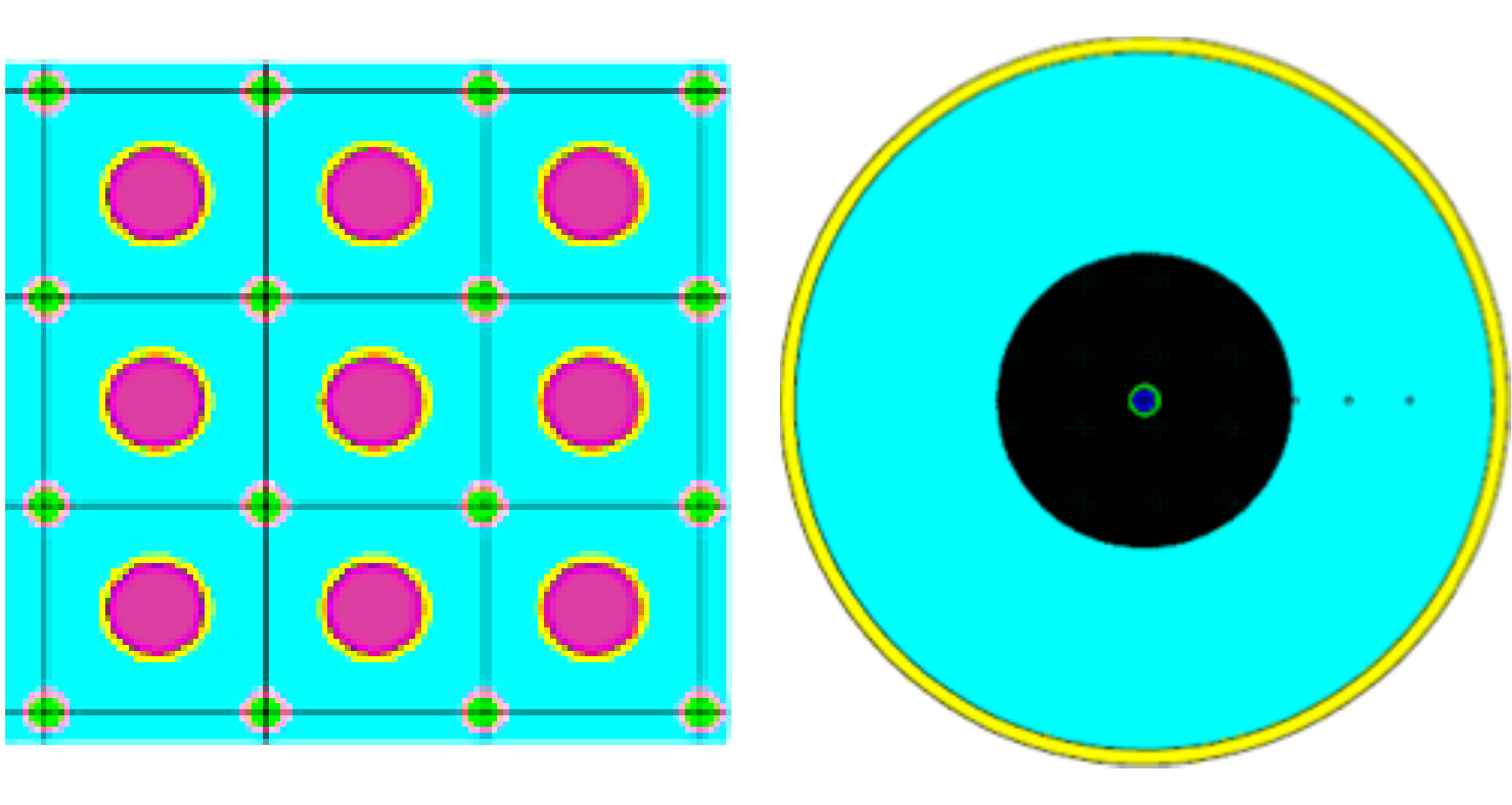}
\caption{On the left (a) the hexahedral core lattice, showing the fuel pins (red), surrounded by a steel cladding (yellow), with the interspersed helium cooling channels (green). On the right (b) the ADS core, showing the central beam pipe (blue with light green contour), the core (black), the lead reflector (cyan) and the steel vessel (yellow).}
\label{Fig:Fig1}
\end{figure}
The fast reactor core was designed (Fig. 1b) as a 1.6x1.6x90 $cm^3$ hexahedral type lattice, containing 0.37 cm radius fuel rods, surrounded by a 0.05 cm thick AISI steel cladding and embedded in a solid lead matrix, for a total of about 3,000 fuel pins. 

Because of the overall cylindrical geometry, fuel pins located at the core boundary can be cut by the reflector inner circumference. The fuel was a mixture of Pu, U and Am oxides (MOX), reported in Table 1. In this configuration, the multiplication coefficient $k_{eff}$ is 0.85 and the total thermal power is 7.9 kW. 

Cooling of the core was obtained by Helium gas flowing in the 0.25 cm diameter steel pipes interspersed between the fuel rods (see Fig. 1a) a cooling scheme that has been shown to be sufficient in similar systems with higher power \cite{RiccoEd}.
\begin{table}[!htb]
\begin{center}
\begin{tabular}{|c|c|}
\hline
Nuclide & Mass Percentage\\ \hline
$^{238}$U & 0.6588 \\ \hline
$^{235}$U & 0.0013 \\ \hline
$^{238}$Pu & 0.0057 \\ \hline
$^{239}$Pu & 0.12197 \\ \hline
$^{240}$Pu & 0.0574 \\ \hline
$^{241}$Pu & 0.0162 \\ \hline
$^{242}$Pu & 0.0176 \\ \hline
$^{241}$Am & 0.00273 \\ \hline
$^{16}$O & 0.1183 \\ \hline
\end{tabular}
\end{center}
\caption{The MOX fuel isotopic vector. The table shows mass percentages for each isotope}
\label{tab:tab1}
\end{table}

This configuration, fairly standard for a lead-based research reactor (Fig. 1b), was taken as the initial model and simulated with MCNP6. After, we introduced some modifications of the reflector and the coolant, studying their effect in terms of thermal power and neutron spectra.

\section{The reflector}
Starting from the initial model, the reflector was modified by inserting between the core and the lead volume a cylindrical shell of graphite acting as intermediate moderator layer (Fig.~2a). We simulated this configuration with various lead/graphite thicknesses $x_{Pb}/x_{Gr}$ (Fig.~2b), keeping the quantity $Q=\rho_{Pb}x_{Pb}+\rho_{Gr}x_{Gr}$ constant and equal to the value for the lead reflector in the initial model (794 gr/cm$^2$) and varying the overall radius of the system correspondingly (the core radius remains the same in all instances). We also simulated the same system with a reflector composed by graphite only (Fig. 2b), without the lead outer shell (in this case clearly the quantity $\rho_{Pb}x_{Pb}+\rho_{Gr}x_{Gr}$ cannot be kept constant).
\begin{figure}[!htb]
\centering
\includegraphics[width=\linewidth]{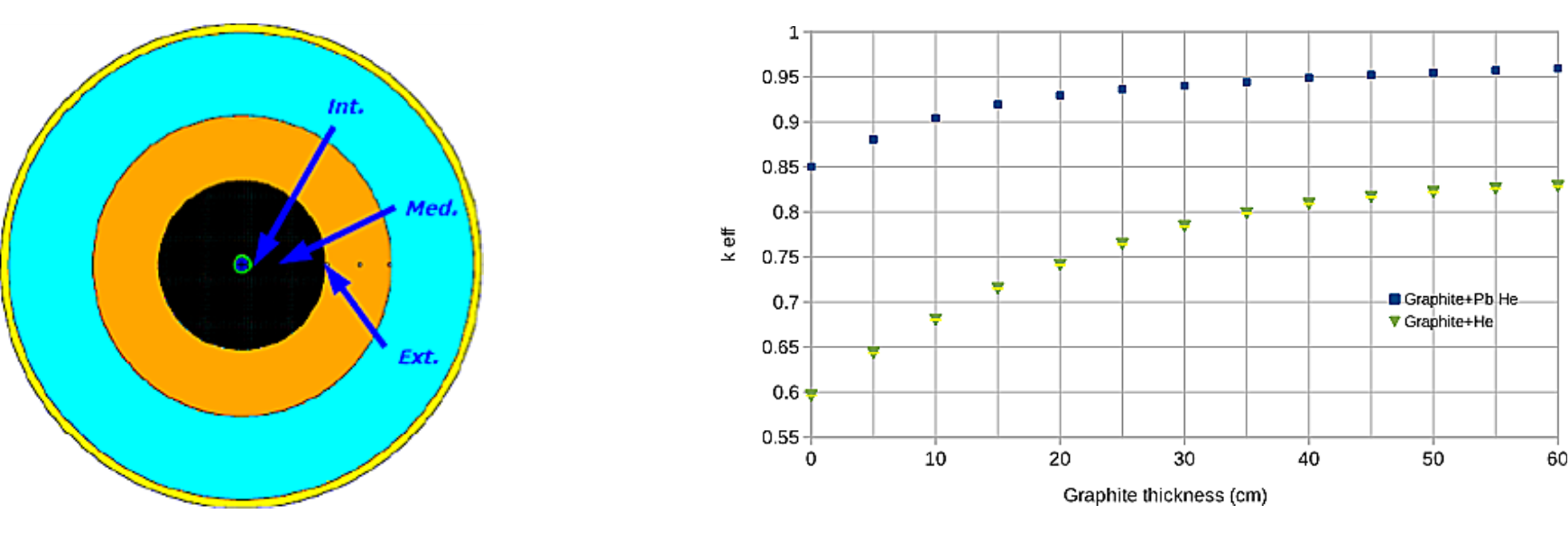}
\caption{On the left (a) the modified reflector comprising a graphite layer (orange) the core (black) and a lead outer shell (light bluel closed by a steel vessel (yellow). The arrows mark the positions of the test pins. On the right (b) the dependence 
of the multiplication factor $k_{eff}$ from the graphite thickness, with and without the outer lead shell.}
\label{Fig:Fig2}
\end{figure}

The core multiplication factor $k_{eff}$, computed using the MCNP6 {\tt kcode}, is plotted in Fig. 2b as a function of the graphite layer thickness with and without the outer lead shell. From the figure two effects appear evident:
\begin{itemize}
\item the insertion of a peripheral graphite moderator layer increases the value of $k_{eff}$, with a significant increase up to about 30-40 cm thickness; we chose 40 cm as the optimal value of the thickness, corresponding to $k_{eff}$=0.95;
\item the presence of the outer lead shell results in higher values of $k_{eff}$ than the case with graphite only, clearly due to the lead reflecting back a significant fraction of the outbound neutrons 
\end{itemize}
As a physical interpretation of these results, it is reasonable to assume that the combination of 
moderator and lead produces both a slowing down and a reflection back into the core of leaking neutrons, thereby producing fissions with higher cross section in the peripheral core region, with a consequent more abundant release of fast fission neutrons. This mechanism should indeed increase the multiplication factor $k_{eff}$, while maintaining the fast character of the neutron energy spectrum in the core, which is indeed what we found and report below.
The energy distribution of neutron fluxes, averaged over the length of selected single fuel pins, were computed both for the initial model and the modified reflector, assuming a 40 cm thick graphite layer, for three positions in the core, reported in fig. 2a: close to the central axis (Fig. 3a), at an intermediate radial position (Fig. 3b) and at the core boundary (Fig. 3c), respectively. 
\begin{figure}[!htbp]
\centering
\includegraphics[width=0.98\linewidth]{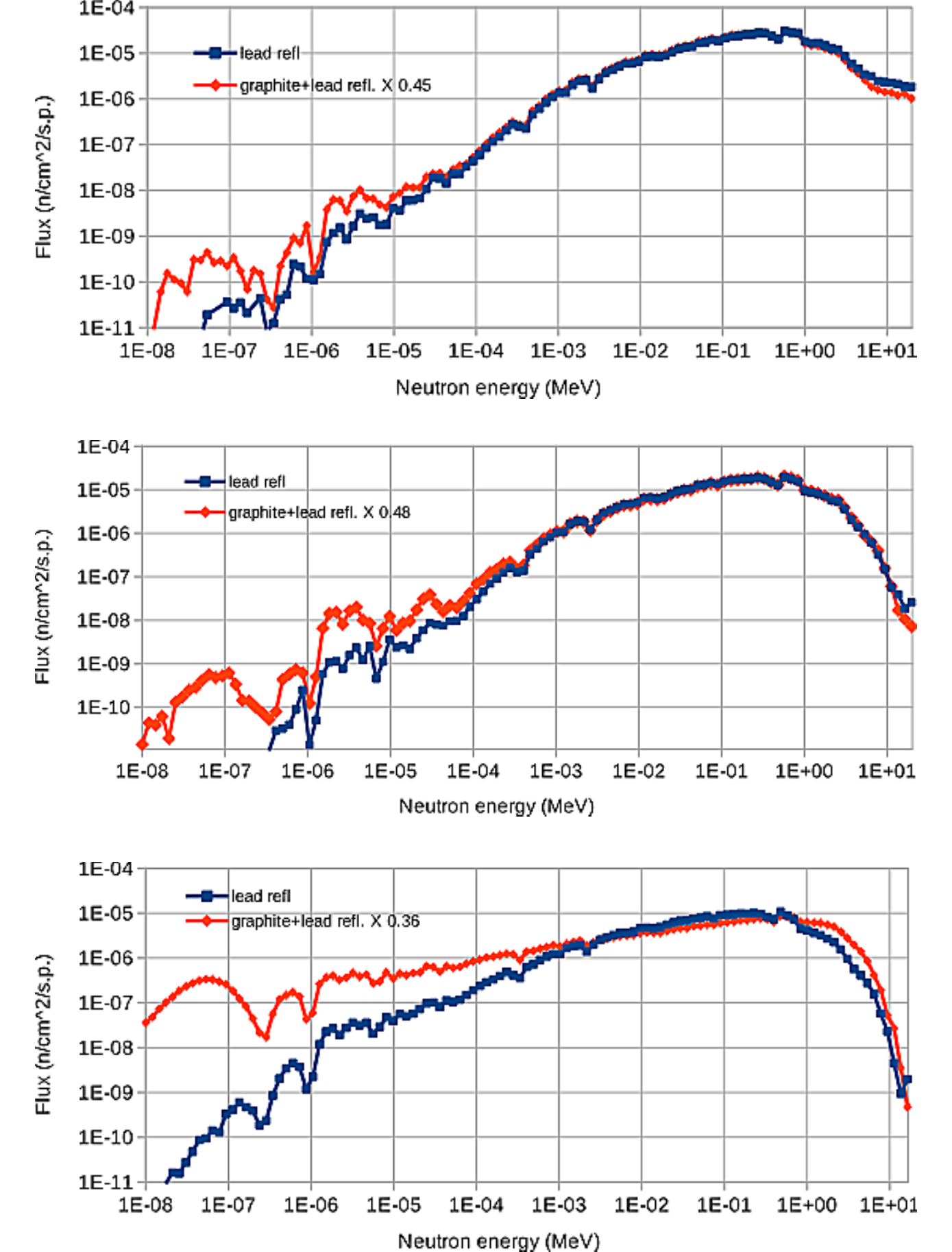}
\caption{The neutron spectrum for the initial model (blue) and for the modified reflector including an intermediate 40 cm graphite layer (red, scaled by the factor indicated in the graphs) at,
from top to bottom, (a) the position close to the central axis, (b) the intermediate radial position and (c) the core boundary.}
\label{Fig:Fig3}
\end{figure}
The insertion of the graphite layer leads to a sizeable increase in the overall neutron flux. Therefore, to allow for a direct comparison between the shapes of the spectra, in Figure 3 we conveniently rescaled - as indicated in the labels - the integrated spectrum by the factor
\begin{equation}
\Frac{\int_0^\infty\,\Phi_{initial}(E) dE}{\int_0^\infty\,\Phi_{Graphite}(E) dE}
\end{equation}
for the case of graphite insertion, so to obtain the same value as for lead alone: this clearly shows that fluxes are systematically higher at all positions, consistently with the larger value for $k_{eff}$. This show also that in the internal and intermediate positions the energy distribution is substantially unchanged, while peripheral pins exhibit a spectrum (Fig. 3c) with an increased soft component (below about 1 keV), a suppressed intermediate component (between about 10 keV and 1 MeV) and an enhanced component above 1 MeV, overall still retaining evident fast characteristics (see Table 2).
\begin{table}[!htb]
\begin{center}
\begin{tabular}{|l|c|c|c|}
\hline
Reflector & Ratio fast/slow & Ratio fast/slow & Ratio fast/slow\\ 
configuration & internal pin & medium pin & external pin\\ \hline
Lead & 2.5 E+5 & 1.7 E+5 & 3.1 E+3\\ \hline
Graphite+lead & 2.9 E+4 & 2.8 E+4 & 2.0 E+1\\ \hline
\end{tabular}
\end{center}
\caption{Ratio between integrated neutron flux above 0.5 MeV (fast) and integrated neutron flux below 1 eV (slow) at the three positions in the core is reported for the initial model (full lead reflector) and the modified model (reflector made of graphite and lead).}
\label{tab:tab2}
\end{table}
The peripheral core behavior is illustrated in more detail in Table 2, where the ratio between the integrated flux above 0.5 MeV and the thermal flux below 1 eV is reported. At the core boundary an enhancement of the fast neutron component is observed for the modified reflector, in the energy range above about 100 keV (Fig. 3c) where fission neutrons are expected to contribute: an explanation that will be confirmed by the thermal power analysis (see Fig. 5) - is that in a narrow region near core boundary a strong enhancement in (thermal) fission activity takes place.
Accordingly, larger fluxes, almost completely thermalized (Fig. 4a), are produced in the graphite reflector, with the thermalization level increasing with the distance from the core (Fig. 4b). In Table 3 the integrated flux up to 1 eV is reported as a percentage of the total flux at three different positions in the graphite part of the reflector in the modified model: 0.5 cm (position C1), 20 cm (C2) and 40 cm (C3) from the core border, respectively.
\begin{figure}[!h]
\centering
\includegraphics[width=0.9\linewidth]{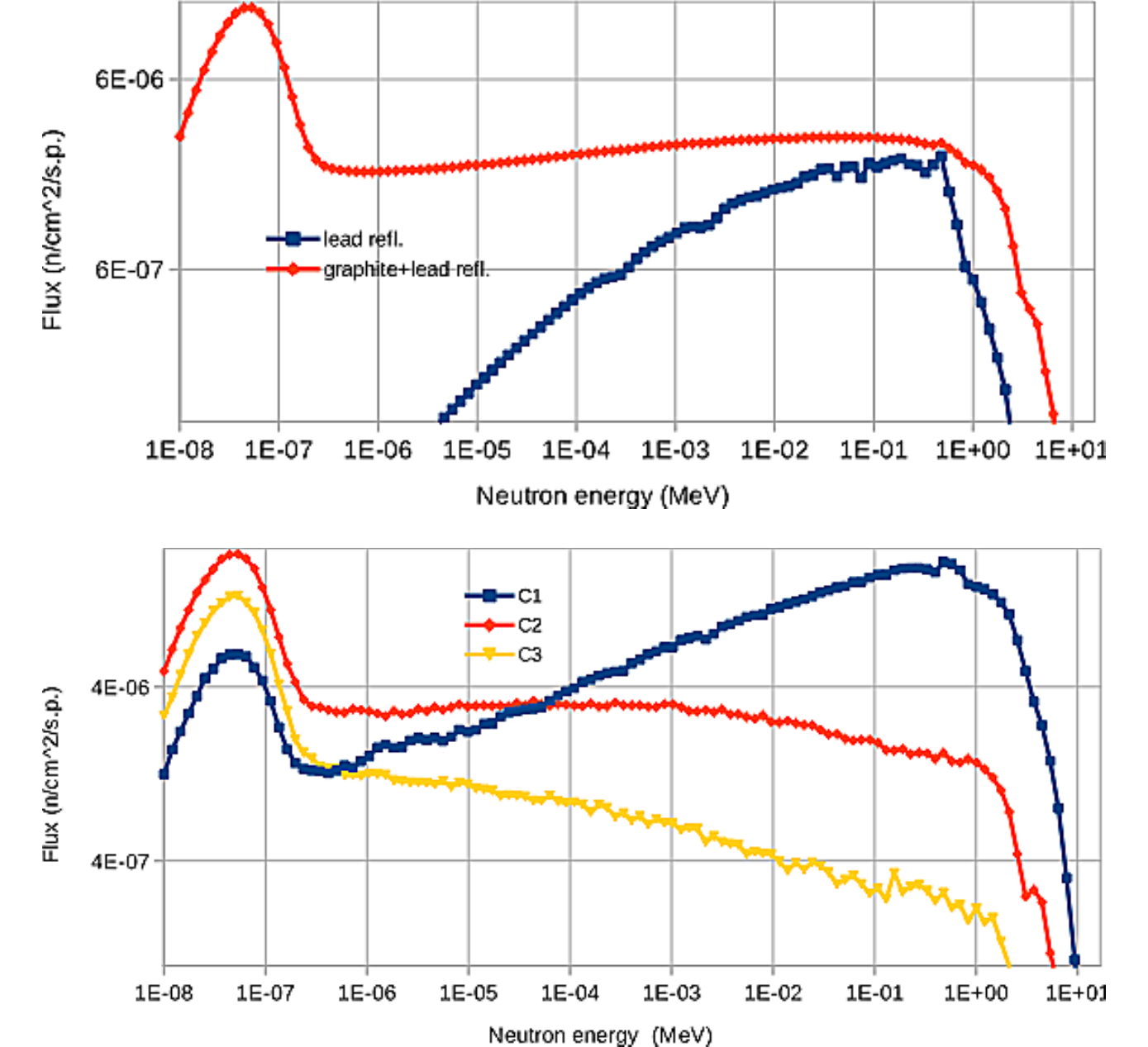}
\caption{(a) Neutron flux averaged in the lead reflector of the initial model (blue) and in the graphite part of the reflector in the modified model (red). (b) Neutron flux at three different positions in the graphite part of the reflector in the modified model: 0.5 cm (C1), 20 cm (C2) and 40 cm (C3) from the core border, respectively.}
\label{Fig:Fig4}
\end{figure}

\begin{table}[!htb]
\begin{center}
\begin{tabular}{|l|c|c|c|}
\hline
Reflector & $\Phi_{< 1eV}$ r=0.5 cm  &  $\Phi_{< 1eV}$ r=20 cm &  $\Phi_{< 1eV}$ r=40 cm\\ 
configuration & (C1) (\%) &  (C2) (\%)  & (C3) (\%) \\ \hline
Lead & 0.09 & 0.43 & 0.82\\ \hline
Graphite+lead & 35 & 78 & 90\\ \hline
\end{tabular}
\end{center}
\caption{The integrated neutron fluxes up to 1 eV at three positions C1, C2 and C3 in the graphite part of the reflector (defined by the indicated distance from the boundary) is reported as a percentage of the total flux, for the initial model (full lead reflector) and the modified model (reflector made of graphite and lead).}
\label{tab:tab3}
\end{table}
As a consequence of this mechanism, a strongly enhanced thermal power is realized along a narrow shell at the core boundary, consistently with the few cm path of thermal neutrons. In fact, the thermal power in individual fuel pins in the case of the layered graphite+lead reflector, plotted in Fig. 5 as a function of the radial distance from the source, shows a pronounced peak, only a few pins wide, at the core boundary. Although in principle advantageous in terms of neutron economy, such a power increase can be an issue in terms of heat removal, even in systems with moderate power, due to the correspondingly large temperature gradient between neighboring border pins.
\begin{figure}[!htb]
\centering
\includegraphics[width=0.98\linewidth]{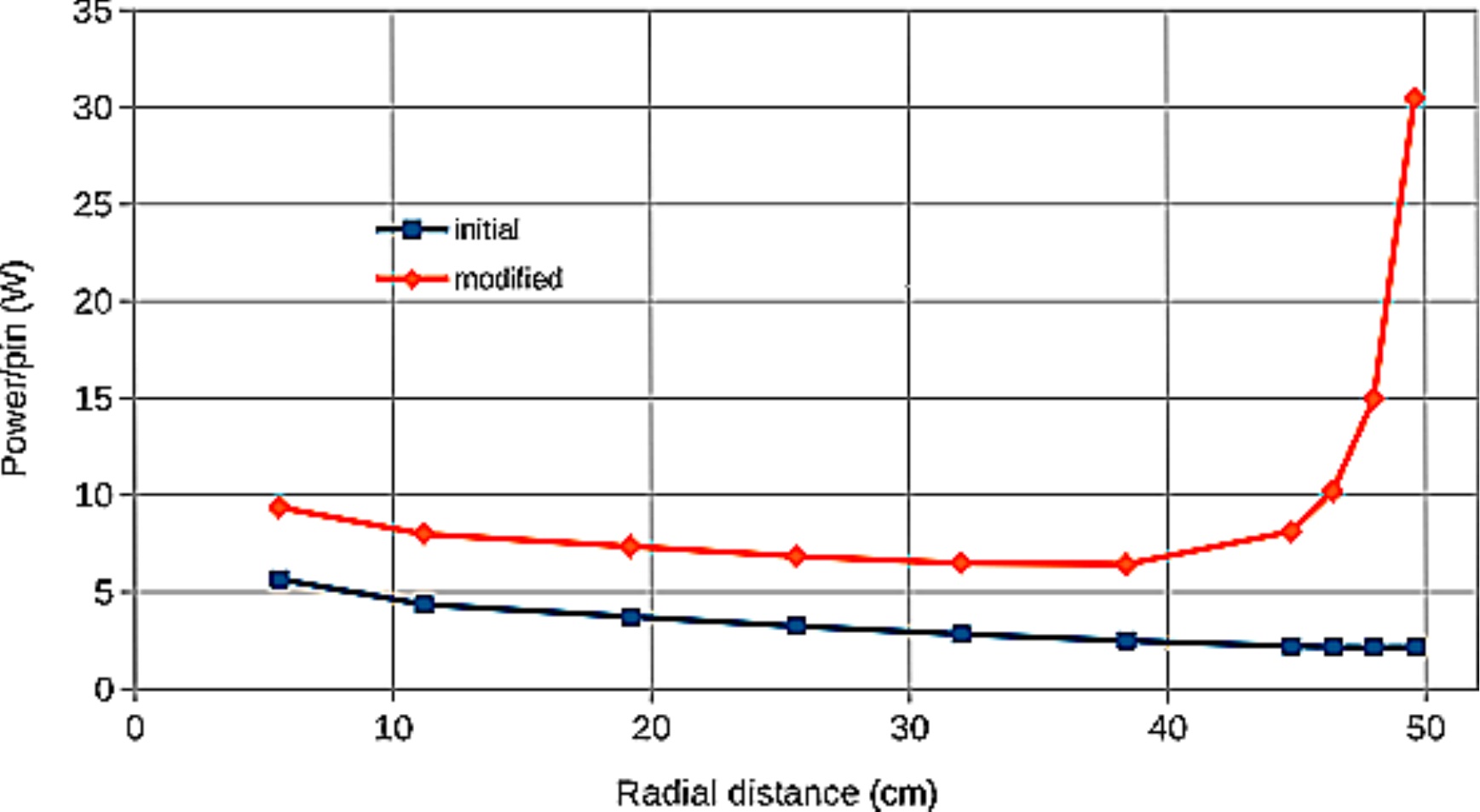}
\caption{Thermal power in individual fuel pins, as a function of the pin number (equivalent to the radial distance from the source), for the initial model (blue) and the modified model (red).}
\label{Fig:Fig5}
\end{figure}
All attempts to reduce this border energy increase necessarily lead, for a given source, to a corresponding reduction of the total produced power, which is mostly concentrated in the peak area shown in Fig. 5. As an example, we also report in Fig. 5 the effect of the insertion, between the 50 cm core and the 40 cm graphite layer, of a further 40 cm lead shell (Fig. 6), reducing the thickness of the external lead reflector to 10 cm, being the latter sufficient to reflect back neutrons that have been significantly slowed down by the first two layers.
\begin{figure}[!htbp]
\centering
\includegraphics[width=0.58\linewidth]{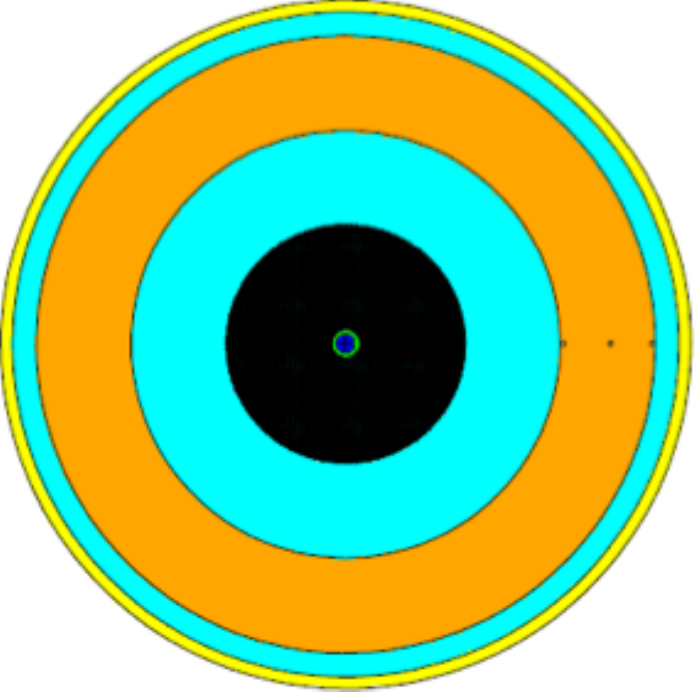}
\caption{ The alternate lead (light blue) and graphite (yellow) layers in the reflector.}
\label{Fig:Fig6}
\end{figure}
Such an inner lead shell will both reflect the fast neutrons from the core and reduce the amount of thermal neutrons that are generated by the graphite in the reflector and that travel back to the core. Fig.~7a shows the effects of such additional inner lead layer on $k_{eff}$.Therefore, the number of thermal neutrons at the periphery of the core is reduced, thereby eliminating the steep power rise. In Fig.~7b, the ratio R between the maximum (peak) and minimum (valley) thermal power per fuel pin along the radial direction (see the red graph in Fig. 5 for the previous case exhibiting valley and peak) is plotted as a function of such inner lead layer, together with the corresponding total power in the core. An almost complete suppression of the power peak at the boundary ($R\simeq 1$) requires a thickness of the additional inner lead layer above 40 cm, with a corresponding strong decrease of the total available power. For the case of an inner lead layer thickness of 40 cm (whose power radial profile is reported in Fig. 5), we obtain keff=0.85 and total power = 8.3 kW.
\begin{figure}[!htbp]
\centering
\includegraphics[width=0.98\linewidth]{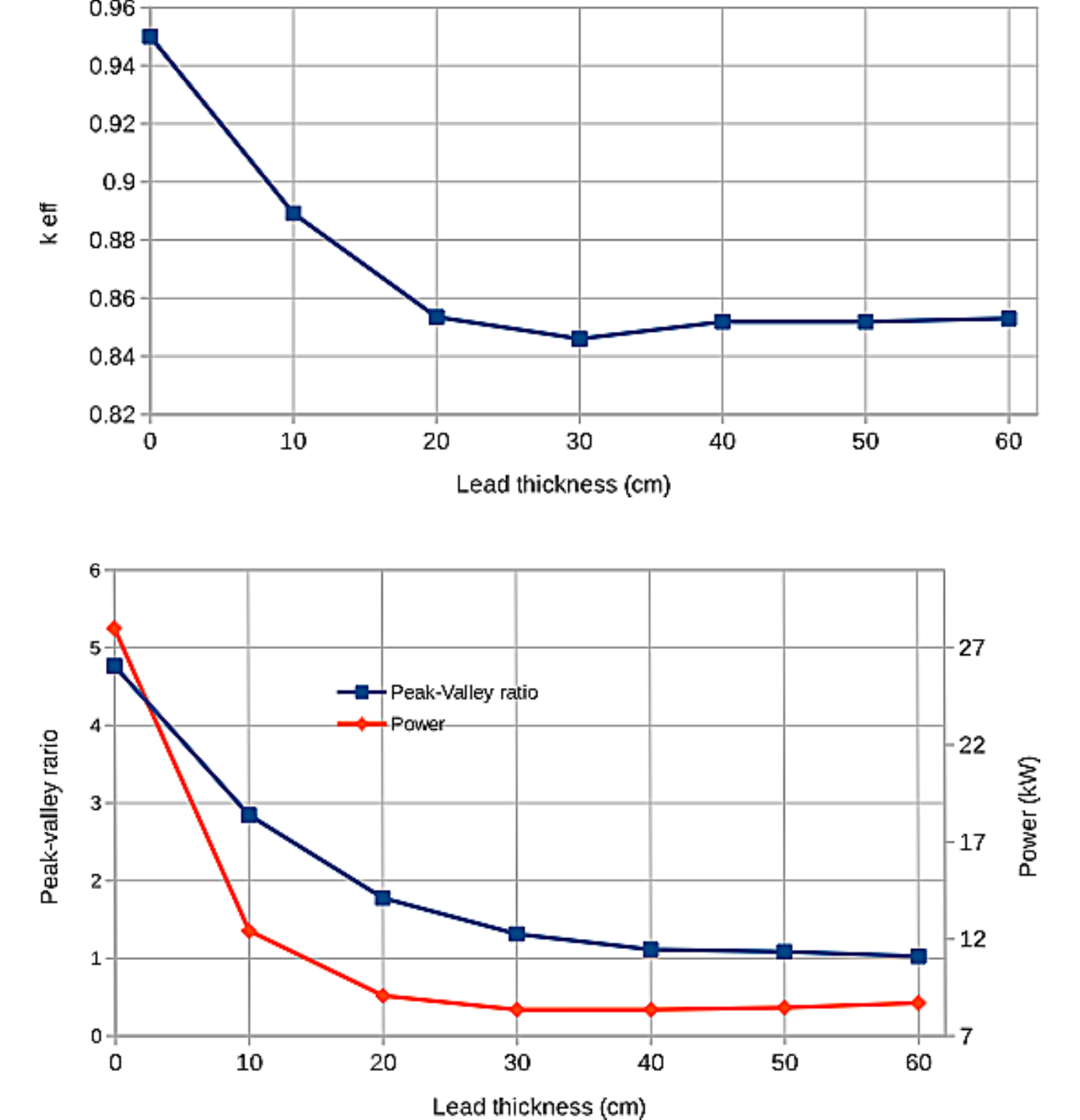}
\caption{From top to bottom (a) $k_{eff}$ as a function of the thickness of the additional inner lead layer between core and graphite shell. (b) Ratio between the maximum (peak) and minimum (valley) power per fuel pin along the radial direction, as a function of the thickness of the additional inner lead layer between core and graphite shell (blue, left scale), together with the corresponding total
power in the core (red, right scale).}
\label{Fig:Fig7}
\end{figure}
While suppressing the border energy excursion, the intermediate 40 cm lead layer restores in the core the pure lead reflector conditions found in the initial model, so that neutron fluxes are practically identical in absolute value and energy dependence to the blue points in Fig. 3a,b,c, except for the most peripheral position at the core boundary (Fig. 8), where a small slow component is evidently present. At the contrary, in the reflector, it strongly affects neutron fluxes in all points (Fig. 9a,b,c), decreasing the absolute values (notice the scale factors in Fig. 9, again calculated as ratios of absolute total fluxes) by about one order of magnitude, but increasing the degree of thermalization (Table 4), with respect to the graphite-lead configuration.
\begin{figure}[!htbp]
\centering
\includegraphics[width=0.98\linewidth]{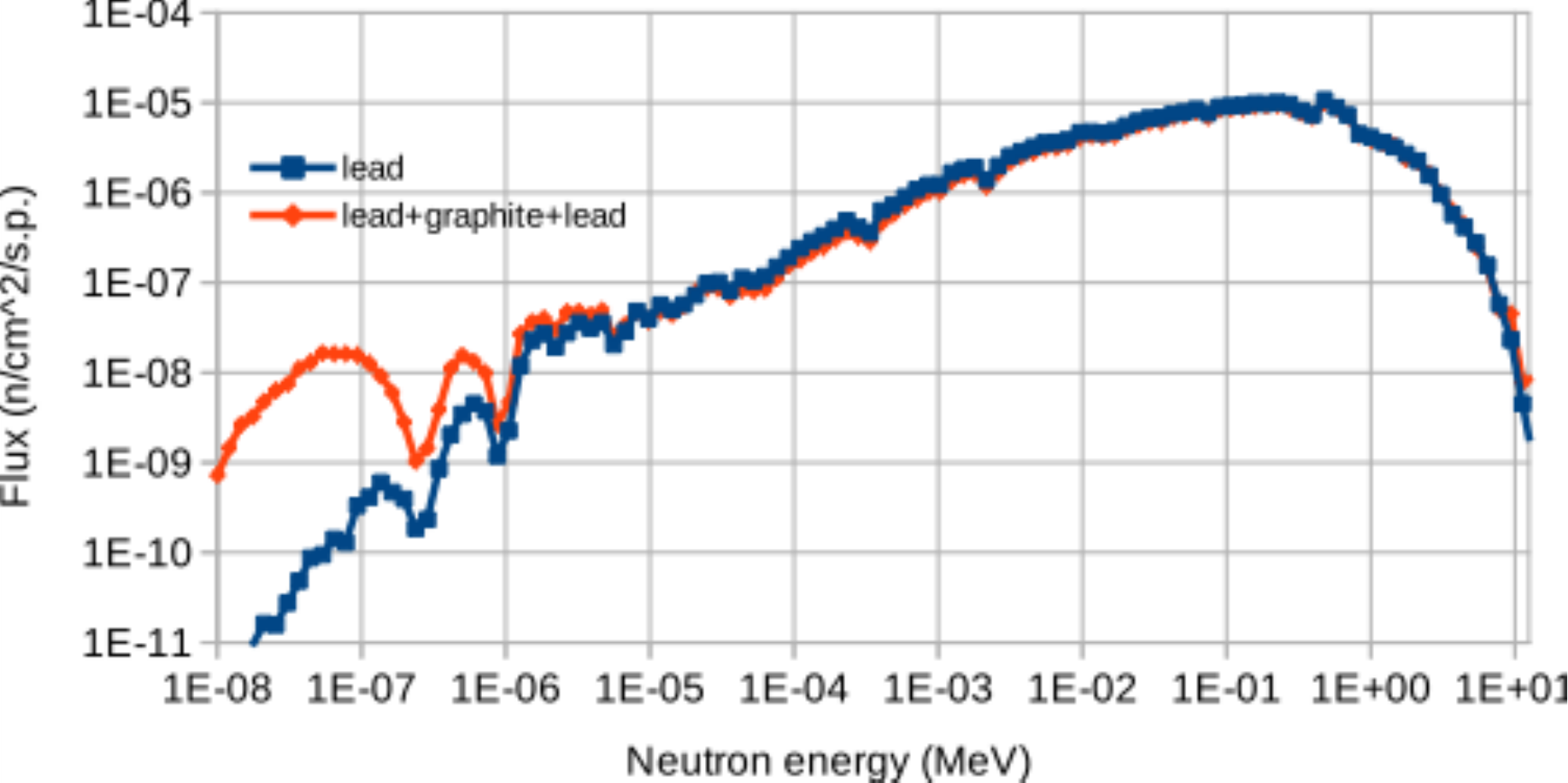}
\caption{ The neutron spectrum for the initial model (blue) and for the modified model with a 40 cm lead- 40 cm graphite-10 cm lead reflector (red) at the core boundary.}
\label{Fig:Fig8}
\end{figure}
In conclusion, the previous discussion suggests the possibility of designing new "hybrid" accelerator driven systems based on a composite reflector, still featuring a fast core, while neutron fluxes with a strong slow/thermal component are obtained in the reflector.
\begin{figure}[!htbp]
\centering
\includegraphics[width=0.85\linewidth]{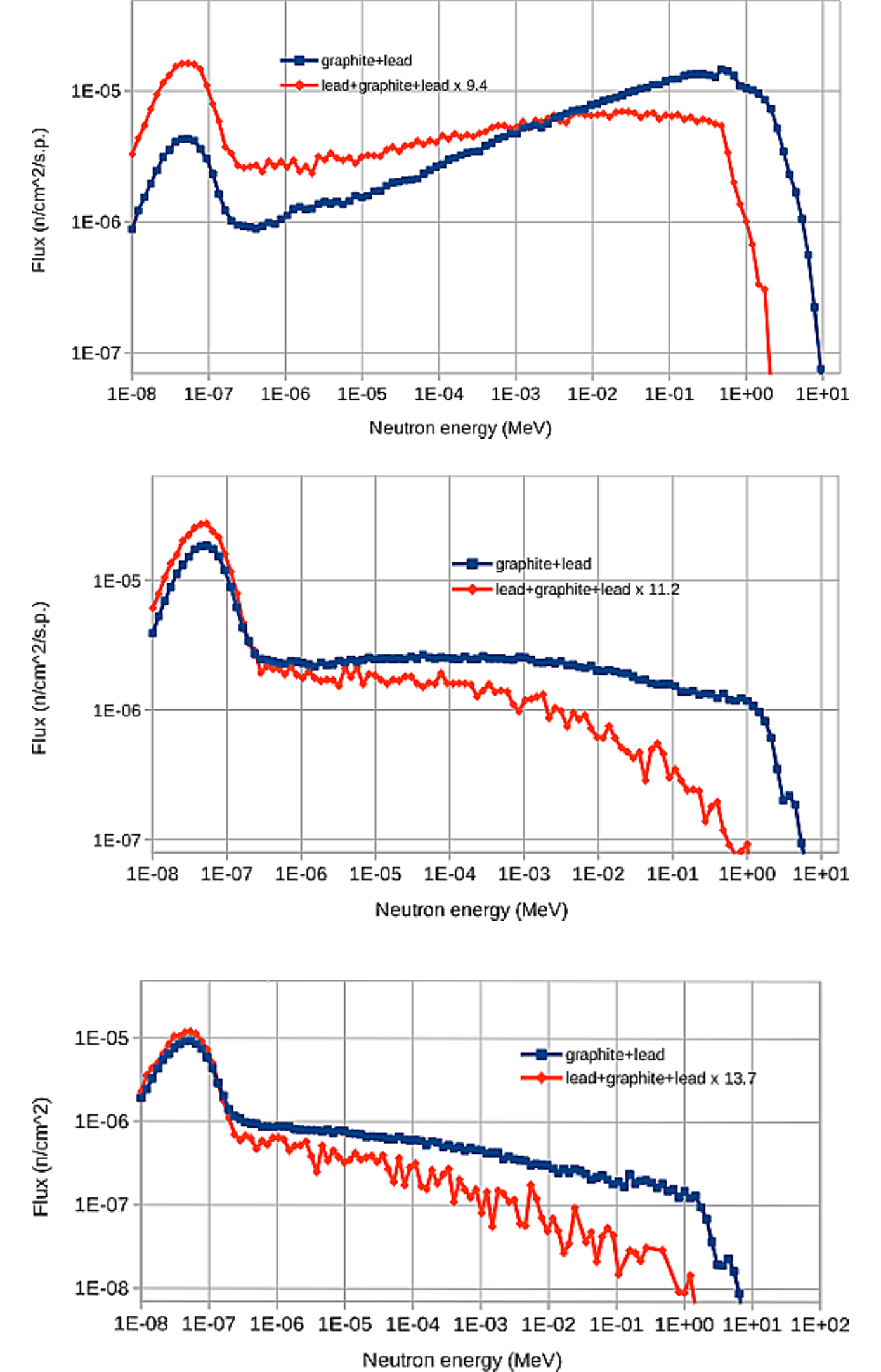}
\caption{Neutron flux in the modified model with subsequent graphite and lead layers in the reflector (blue) and in the further modified model with a 40 cm lead-40 cm graphite-10 cm lead reflector (red) at three different positions in the reflector: from top to bottom 0.5 cm (a), 20 cm (b) and 40 cm (c) from the core border, respectively, for the graphite-lead model; 40.5 cm (a), 60 cm (b) and 80 cm (c) from the core border, respectively, for the lead-graphite-lead model.}
\label{Fig:Fig9}
\end{figure}

\begin{table}[!htb]
\begin{center}
\begin{tabular}{|l|c|c|c|}
\hline
Reflector & $\Phi_{< 1eV}$ r=0.5 cm  &  $\Phi_{< 1eV}$ r=20 cm &  $\Phi_{< 1eV}$ r=40 cm\\ 
configuration &  (\%) &   (\%)  &  (\%) \\ \hline
Graphite+lead & 10 & 57 & 74\\ \hline
Reflector & $\Phi_{< 1eV}$ r=40.5 cm  &  $\Phi_{< 1eV}$ r=60 cm &  $\Phi_{< 1eV}$ r=80 cm\\ 
configuration &  (\%) &   (\%)  &  (\%) \\ \hline
Lead+graphite & 35 & 78 & 90\\ 
+lead & & &\\
\hline
\end{tabular}
\end{center}
\caption{The integrated neutron fluxes up to 1 eV at three positions in the graphite part of the reflector (defined by the indicated distance from the boundary) is reported as a percentage of the total flux, for the graphite-lead model and the 40 cm lead-40 cm graphite-10 cm lead model.}
\label{tab:tab4}
\end{table}
\section{The coolant}
In the moderator/reflector configuration, a smooth thermal power radial distribution can be obtained in the border region at the expenses of the available total power (Fig. 6): for example, in the geometry described in Section 2, the 40 cm inner lead layer inserted between the core and the graphite shell reduces the $k_{eff}$ from 0.95 (Fig. 1b) to 0.84. However, if a small surplus of low energy neutrons could be produced somehow uniformly in the core, we might be able to bring back $k_{eff}$ to the previous level by exploiting the enhancement in fission rates produced all over the core, without introducing discontinuities across the core volume and - hopefully - without perturbing the fast character of the neutron spectrum in the core.
To explore this possibility, we modified the initial model of Section 1, featuring a pure lead reflector, by simply replacing the He coolant (Fig. 1a) with light water. The comparison between the two configurations shows an increase of keff from 0.85 to 0.92 and a very similar energy distribution of scaled neutron fluxes (Fig. 10a,b,c).
\begin{figure}[!htbp]
\centering
\includegraphics[width=0.98\linewidth]{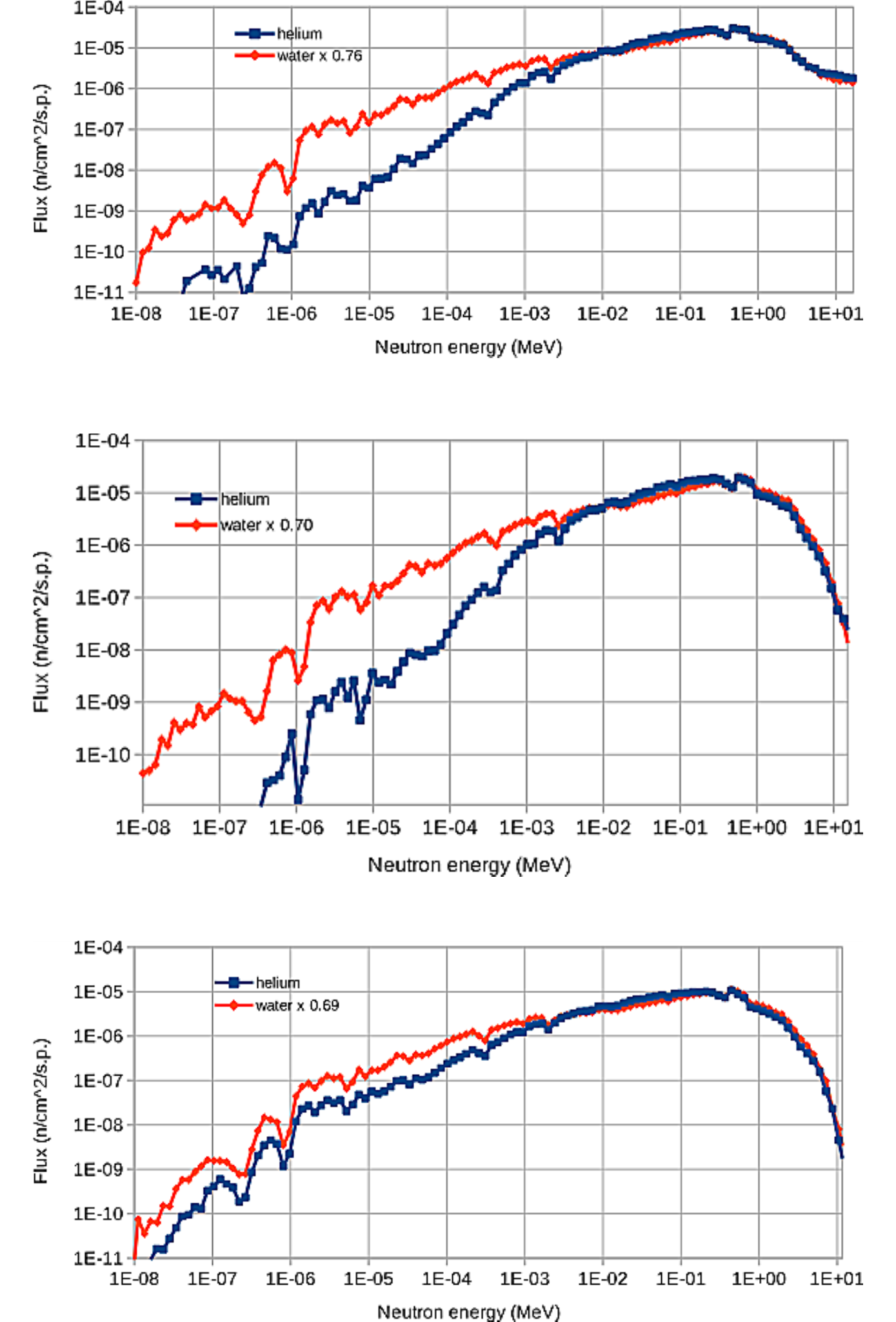}
\caption{From top to bottom (a) The spectrum at the position close to the central axis (a), at the intermediate radial position (b) and at the core boundary (c), for the initial model (blue) and for the version where light water has replaced helium as coolant (red, where 0.76, 0.7 and 0.69 represent scale factors).}
\label{Fig:Fig10}
\end{figure}

In Table 5, the relative weight of integrated fast ($>0.5$ MeV) and slow ($<1$ eV) flux components are reported for the two coolants in three different positions.
\begin{table}[!htb]
\begin{center}
\begin{tabular}{|l|c|c|c|c|c|c|c|}
\hline
Reflector & Coolant & Ratio & Ratio & Ratio & $\Phi_{< 1 eV}$ & $\Phi_{< 1 eV}$ & $\Phi_{< 1 eV}$ \\
Config. &       & fast/slow & fast/slow & fast/slow & internal & medium & external \\
               &               & internal &  medium & external & pin (\%) & pin (\%) & pin (\%) \\
               &               &  pin       &   pin       &  pin          &   & &\\ \hline
 Pb    & He & 2.5E+5 & 1.7E+5 & 3.1E+3 & 4.E-5 & 1.13 E-4 & 7.82E-3\\ \hline
 Pb-C-Pb & He & 1.8E+5 & 1.1E+5 & 2.7E+2 & 2.1E-4 & 2.8E-4 & 9.0E-2\\ \hline
 Pb    & H$_2$O & 3.8E+3 & 2.8E+3 & 1.0E+3 & 8.2E-3 & 2.6E-2 & 2.4E-2\\ \hline
\end{tabular}
\end{center}
\caption{Ratio between integrated neutron flux above 0.5 MeV (fast) and integrated neutron flux below 1 eV (slow), together with relative weight of integrated slow ($<1$ eV) flux components, reported for the three cases considered (initial model, layered reflector and single-layer reflector with water as coolant) at the three above mentioned different positions in the core.}
\label{tab:tab5}
\end{table}
The effect of water as coolant results in a limited increase, with negligible thermalization, of the slow flux component without significantly affecting the fast character of the core spectrum.
An important aspect, following the discussion of Section 2, is the radial distribution of the deposited energy in the core, reported in Fig. 11: the helium and water coolants produce a very similar smooth behavior without discontinuities, the nominal power being higher in the water case, consistent with the higher multiplication factor.
\begin{figure}[!htbp]
\centering
\includegraphics[width=0.98\linewidth]{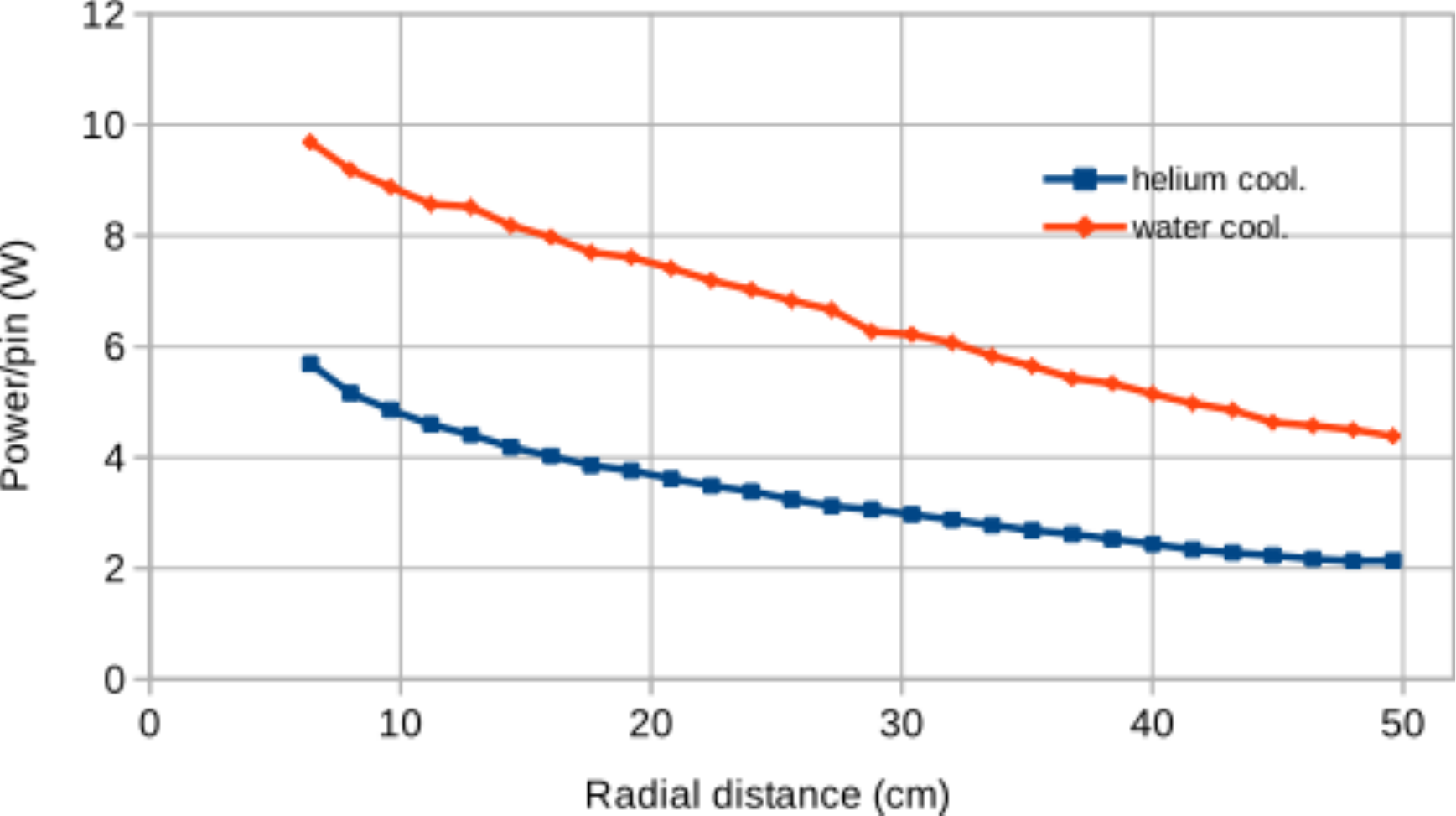}
\caption{Thermal power per source neutron in individual fuel pins, as a function of the pin number (equivalent to the radial distance from the source), for the initial model Helium cooled and water cooled (red).}
\label{Fig:Fig11}
\end{figure}

\section{Conclusions}
We studied the influence of specific choices of reflector materials and core coolant on the characteristics of and accelerator driven system. By using a simplified model of a fast ADS we performed extensive simulations with the MCNP6 code to investigate the behavior of the system in terms of multiplication coefficient $k_{eff}$, thermal power and absolute neutron spectra. First, we studied the effects of replacing pure lead with a layered structure of graphite and lead as reflector around the core. We found that, by appropriately choosing position and thickness of the graphite and lead shells, it is possible to obtain a "hybrid" system where the neutron spectrum in the core still exhibits a fast character, while the neutron spectrum in the graphite shell is considerably softer, with a marked thermal character in the most peripheral positions. In the case that the graphite shell was directly facing the core, we observed instead a steep spatial power increase at the core boundary, which may be an issue in terms of a too high power gradient across the fuel pins and other materials. This unwanted side effect of local power increase can be circumvented by inserting a second lead shell between the core and the moderating graphite layer, thereby moving the graphite away from the close vicinity to the fuel, although at the expenses of decreasing the value of keff and therefore the total power of the system. To bring $k_{eff}$ and the power back to the desired level, we explored the option of replacing the helium gas with light water as coolant. We found that this is possible without significantly perturbing the fast character of the system and without introducing spatial power excursions in any place within the core.
The total power for each discussed configuration with the same external neutron source intensity ( $8. 10^{14}$ n/sec) is reported in Table 6:
\begin{table}[h]
\begin{center}
\begin{tabular}{|l|c|c|c|c|}
\hline
Reflector & Lead & Lead & Graphite/Lead & Lead/Graphite/Lead \\ \hline
Coolant   & Helium & Water & Helium & Helium \\ \hline
Power (kW) & 7.9 & 17.7 & 27.8 & 8.3 \\ \hline
\end{tabular}
\end{center}
\caption{Total power in the different reflector and coolant configurations.}
\label{tab:tab6}
\end{table}

In conclusion, a "hybrid" ADS where neutron spectra are fast in the core and softer, even thermal, outside of the core seems possible by inserting a graphite shell of appropriate thickness in the lead reflector. By replacing helium with light water in the cooling channels interspersed in the lead matrix hosting the fuel pins, it is possible to somewhat boost the thermal power of the system. Such a "hybrid" system may offer the advantage of providing both fast, slow and thermal neutrons that may be used for a variety of experiments and studies on waste transmutation, material activation and fuel evolution under irradiation. The application of the study presented here to a more realistic research reactor design will be reported in a forthcoming paper.

\section*{Acknowledgments}
This work is partially supported by the the European Atomic Energy Community's (Euratom) Seventh Framework Program FP7/2007-2011 under the Project CHANDA (Grant No. 605203), by the Centro Fermi under the project "Intrinsically Safe Systems" and by the Istituto Nazionale di Fisica Nucleare under the "INFN\_E" strategic project.


\newpage
\section*{References}
\begin{thebibliography}{9}
\bibitem{GenIV}Technology Roadmap Update for Generation IV Nuclear Energy Systems (2014), OECD Nuclear Energy Agency for the Generation IV International forum, available on line https://www.gen-4.org/gif/upload/docs/application/pdf/2014-03/gif-tru2014.pdf

\bibitem{JHR}G. Bignan, C. Colin, J. Pierre, C. Blandin, C. Gonnier, M. Auclair, F. Rozenblum, "The Jules Horowitz Reactor Research Project: A New High Performance Material Testing Reactor Working as an International User Facility - First Developments to Address R\&D on Material", EPJ Web of Conferences 115 (2016)01003.

\bibitem{Nifene}H. Nifenecker, S. David, J.M. Loiseaux, O. Meplan, "Review: Basics of accelerator driven subcritical reactors", Nucl. Instr. Meth. A463 (2001)428-467]

\bibitem{RiccoEd}"An intrinsically safe facility for forefront research and training on nuclear technologies", edited by G. Ricco, Eur. Phys. J. Plus (2014) 129

\bibitem{Misha}Osipenko, M. et al., "Measurement of neutron yield by a 62 MeV proton beam on a thick Beryllium target", Nucl. Inst. Methods A 723, 8 (2013).
\end{thebibliography}
\end{document}